\documentstyle[multicol,pre,aps,epsf,amsmath]{revtex}
\begin{document}

\draft

\title{Efficient method for simulating quantum electron dynamics\\ under the time dependent Kohn-Sham equation}

\author{Naoki Watanabe\, and\, Masaru Tsukada}

\address{Department of Physics,%
  Graduate School of Science, University of Tokyo,
  7-3-1 Hongo, 113-0033 Bunkyo-ku, Tokyo, Japan}

\date{to be published from Physical Review E, 2002.}

\maketitle

\begin{abstract}
A numerical scheme for solving the time-evolution of wave functions under
the time dependent Kohn-Sham equation has been developed. Since the effective
Hamiltonian depends on the wave functions, the wave functions and the effective
Hamiltonian should evolve consistently with each other. For this purpose,
a self-consistent loop is required at every time-step
for solving the time-evolution numerically, which is computationally expensive.
However, in this paper, we develop a different approach expressing a formal
solution of the TD-KS equation, and prove that it is possible
to solve the TD-KS equation efficiently and accurately by means of a simple 
numerical scheme without the use of any self-consistent loops.
\end{abstract}


\begin{multicols}{2}
\narrowtext

\section{Introduction}

Since the innovative work on the density functional theory (DFT)
\cite{Hohenberg1964} and the Kohn-Sham equation \cite{KohnSham1965},
many kinds of static or adiabatic quantum electronic phenomena have been
investigated based on first principles. As an extension of the DFT to
non-adiabatic dynamical phenomena, the time-dependent density functional theory
(TD-DFT) has been developed\cite{Runge1984,Gross1995}. By using the TD-DFT,
some excitation phenomena have been analyzed more accurately than by using
the DFT\cite{Theilhaber1992}.
However, the formulation of the TD-DFT is too complicated to solve
the wave functions numerically in order to see electron dynamics directly.
So a considerable approximate formula called the TD-Kohn-Sham (TD-KS) equation
has been applied for the numerical simulations
\cite{Sugino1999,Iwata2000}.

The difficulty in numerically solving the TD-KS equation is the treatment
of the density-dependent Hamiltonian. The wave functions and the Hamiltonian
should always be self-consistent with each other.
A fourth order self-consistent iterative scheme was proposed by
O. Sugino and Y. Miyamoto\cite{Sugino1999}. However, the use of a SCF-loop
at every time-step is computationally expensive.

In this paper, we propose a new formalism for the numerical solution of
the TD-KS equation. Based it on, we prove that a simple formula without
SCF-loops can solve the TD-KS equation with sufficient accuracy.
We find that computational techniques\cite{Watanabe2000a,Watanabe2000b}
previously developed by us for the one-electron TD-Schr\"odinger equation in real
space and real time are also useful for the TD-KS equation.

\section{Conventional method}

The TD-KS equation is a mean field approach used for describing the
time-evolution of the electron density $\rho$ via one-electron wave
functions $\psi_n$ under an effective Hamiltonian ${\cal H}$,
\begin{gather}
  {\rm i} \frac{\partial \psi_n(t)}{\partial t}
=
  {\cal H}[\rho,t]\, \psi_n(t) \ ;\quad
  {\cal H}[\rho,t]
=
  - \frac{\triangle}{2} + V[\rho,t]\ , \label{2-1} \\
  V[\rho,t] = V_{\rm int}[\rho] + V_{\rm ext}(t) \ , \quad
  \rho(t) = \sum_{n=1}^N |\psi_n(t)|^2 \ . \notag
\end{gather}
Here, $V[\rho,t]$ is an effective potential which represents the internal
mutual interactions $V_{\rm int}[\rho]$ and the external time-dependent
potential $V_{\rm ext}(t)$.
Throughout this paper, we use the atomic unit $\hbar=1,\,m=1,\,e=1$
for equations and values.

Due to the time-dependence of the Hamiltonian,
the solution of the TD-KS equation can be formally 
expressed in terms of a time-ordering exponential operator:
\begin{equation}
  \psi_n(t)
=
  {\cal T}
  \exp\Bigl[
  {-{\rm i}
  \int_{0}^{t} {\rm d}t^\prime\,
  {\cal H}[\rho,t^\prime] \Bigr]}\,
  \psi_n(0) \ .
  \label{2-2}
\end{equation}

There are many numerical methods for computing Eq.~(\ref{2-2}).
The simplest method discretizes the elapsed time $t$ into small time slices
$\Delta{t}$, and approximates Eq.~(\ref{2-2}) as
\begin{equation}
  \psi_n(t+\Delta{t})
\approx
  \exp{\Bigl[ -{\rm i} \Delta{t} \, {\cal H}[\rho,t] \Bigr]}\,
  \psi_n(t) \ ,
  \label{2-3}
\end{equation}
and it is computed using the Runge-Kutta method,
or by the split operator technique:
%
\begin{multline}%
  \psi_n(t+\Delta{t})
\sim
  \exp{\Bigl[ \frac{{\rm i}\Delta{t}}{2} \frac{\triangle}{2} \Bigr]}
  \exp{\Bigl[ \frac{\Delta{t}}{{\rm i}} V[\rho,t] \Bigr]} \\%
  \exp{\Bigl[ \frac{{\rm i}\Delta{t}}{2} \frac{\triangle}{2} \Bigr]}
  \psi_n(t) \ .
  \label{2-4}
\end{multline}%
%
However, this is not sufficiently accurate, because it
ignores the time dependence of the Hamiltonian during the small time slice,
while the splitting reduces accuracy to an even lower level.

Another well-known computational method for Eq.~(\ref{2-2})
uses a Hamiltonian in the middle of the steps,
\begin{equation}
  \psi_n(t+\Delta{t})
\simeq
  \exp{\Bigl[ -{\rm i} \Delta{t}\, {\cal H}[\rho,t+\frac{\Delta{t}}{2}] \Bigr]}\,
  \psi_n(t) \ .
  \label{2-5}
\end{equation}
Eq.~(\ref{2-5}) is also computed by the split operator technique:
%
\begin{multline}%
  \psi_n(t+\Delta{t})
\sim
  \exp{\Bigl[ \frac{{\rm i}\Delta{t}}{2} \frac{\triangle}{2} \Bigr]}
  \exp{\Bigl[ \frac{\Delta{t}}{{\rm i}} V[\rho,t+\frac{\Delta{t}}{2}] \Bigr]}\\%
  \exp{\Bigl[ \frac{{\rm i}\Delta{t}}{2} \frac{\triangle}{2} \Bigr]}
  \psi_n(t) \ .
  \label{2-6}
\end{multline}%
%
Here, $V[\rho,t+\Delta{t}/2]$ is estimated from an interpolation
between $V[\rho,t]$ and $V[\rho,t+\Delta{t}]$. Therefore, they
have to be solved by a self-consistent loop.
This scheme is accurate enough; however, it is computationally
expensive to perform the SCF-loop at every time-step.\\[-10mm]

\section{Formulation}

To avoid the use of a SCF-loop, we first express the time-evolution
of wave functions using a Taylor development in exponential form as
\begin{equation}
  \psi_n(t+\Delta{t})
=
  \sum_{k=0}^{\infty} \frac{\Delta{t}^k}{k!} \frac{\partial ^k}{\partial t^k}
  \psi_n(t)
=
  \exp{\Bigl[
    \Delta{t} \frac{\partial }{\partial t}
  \Bigr]}\, \psi_n(t)\ .
  \label{3-1}
\end{equation}
We consider a quantity $f(\{\psi\},\{\psi^\ast\},t)$
which depends on wave functions $\psi$ and time $t$ explicitly.
The time-derivative of this quantity is expanded by the chain rule,
\begin{equation}
  \frac{\partial f}{\partial t}
=
    \frac{\partial \psi     }{\partial t} \cdot
    \frac{\delta f}{\delta \psi}
  + \frac{\partial \psi^\ast}{\partial t} \cdot
    \frac{\delta f}{\delta \psi^\ast}
  + \frac{\partial f}{\partial t_{\rm ex}} \ .
  \label{3-2}
\end{equation}
Here, we have used the following notation,
\begin{equation}
    \frac{\partial \psi     }{\partial t} \cdot
    \frac{\delta f}{\delta \psi}
\equiv
    \sum_{m=1}^N \int {\rm d}{\bf r} 
    \frac{\partial \psi_m({\bf r})     }{\partial t}\,
    \frac{\partial f}{\partial \psi_m({\bf r})} \ ,
  \label{3-3}
\end{equation}
and ${\partial}/{\partial t_{\rm ex}}$ means
an explicate-time-derivative operator,
which operates only explicitly-time-dependent quantities.

By substituting the TD-KS equation (\ref{2-1}) into Eq.~(\ref{3-2}),
the time-differential is generally expressed as
\begin{equation}
  {\rm i} \frac{\partial }{\partial t}
=
  ({\cal H}[\rho,t] \psi)      \cdot \frac{\delta}{\delta \psi}
- ({\cal H}[\rho,t] \psi)^\ast \cdot \frac{\delta}{\delta \psi^\ast}
+ {\rm i} \frac{\partial}{\partial t_{\rm ex}} \ .
  \label{3-4}
\end{equation}
For example, it operates a wave function $\psi_n$ as
\begin{align}
  {\rm i} \frac{\partial \psi_n}{\partial t}
&=
  ({\cal H}[\rho,t] \psi)      \cdot \frac{\delta\psi_n}{\delta \psi}
- ({\cal H}[\rho,t] \psi)^\ast \cdot \frac{\delta\psi_n}{\delta \psi^\ast}
+ {\rm i} \frac{\partial\psi_n}{\partial t_{\rm ex}}  \notag\\
&=
  {\cal H}[\rho,t] \psi_n \ ,
  \label{3-5}
\end{align}
because $\psi_n$ does not depend on $\psi_m^\ast$ and $t$ explicitly.

Another example regards density $\rho$,
\begin{align}
  {\rm i} \frac{\partial \rho}{\partial t}
&=
  ({\cal H}[\rho,t] \psi)      \cdot \frac{\delta\rho}{\delta \psi}
- ({\cal H}[\rho,t] \psi)^\ast \cdot \frac{\delta\rho}{\delta \psi^\ast}
+ {\rm i} \frac{\partial\rho}{\partial t_{\rm ex}}  \notag\\
&=
  \sum_m
  ({\cal H}[\rho,t] \psi_m)      \psi_m^\ast
- ({\cal H}[\rho,t] \psi_m)^\ast \psi_m \ ,
  \label{3-6}
\end{align}
because $\rho$ also does not depend on $t$ explicitly.

By substituting Eq.~(\ref{3-4}) into Eq.~(\ref{3-1}), we can formally
write the solution without employing the time-ordering operator as
%
\begin{multline}%
  \psi_n(t+\Delta{t})
=
  \exp \frac{\Delta{t}}{\rm i} \Bigl[
    ({\cal H}[\rho,t] \psi)      \cdot \frac{\delta}{\delta \psi} \\%
  - ({\cal H}[\rho,t] \psi)^\ast \cdot \frac{\delta}{\delta \psi^\ast}
  + {\rm i} \frac{\partial}{\partial t_{\rm ex}}
  \Bigr] \, \psi_n(t) \ .
  \label{3-7}
\end{multline}%
%
However, it does not describe the algorithm of computations.
To show the way of computation of Eq.(\ref{3-7}), we decompose the exponential
operator as,
\begin{multline}
  \psi_n(t+\Delta{t}) \simeq
  \exp\Bigl[
    \frac{\Delta{t}}{2} \frac{\partial}{\partial t_{\rm ex}}
  \Bigr] \\%
  \exp \frac{{\rm i}\Delta{t}}{4} \Bigl[
    (\triangle\psi)      \cdot \frac{\delta}{\delta \psi}
  - (\triangle\psi)^\ast \cdot \frac{\delta}{\delta \psi^\ast}
  \Bigr] \\
  \exp \frac{\Delta{t}}{{\rm i}} \Bigl[
    (V[\rho,t]\psi)      \cdot \frac{\delta}{\delta \psi}
  - (V[\rho,t]\psi)^\ast \cdot \frac{\delta}{\delta \psi^\ast}
  \Bigr] \\
  \exp \frac{{\rm i}\Delta{t}}{4} \Bigl[
    (\triangle\psi)      \cdot \frac{\delta}{\delta \psi}
  - (\triangle\psi)^\ast \cdot \frac{\delta}{\delta \psi^\ast}
  \Bigr] \\%
  \exp\Bigl[
    \frac{\Delta{t}}{2} \frac{\partial}{\partial t_{\rm ex}}
  \Bigr]
  \psi_n(t) \ .
  \label{3-8}
\end{multline}
Equation (\ref{3-8}) is correct up to the second-order of $\Delta{t}$.

To clarify the meaning of the exponential operator which contains
the Laplacian appearing in Eq.~(\ref{3-8}),
we expand it in a Taylor development as
\begin{multline}
  \exp \frac{{\rm i}\Delta{t}}{4} \Bigl[
    (\triangle\psi)      \cdot \frac{\delta}{\delta \psi}
  - (\triangle\psi)^\ast \cdot \frac{\delta}{\delta \psi^\ast}
  \Bigr] \psi_n \\%
  =
  \sum_{k=0}^{\infty} \frac{({\rm i}\Delta{t})^k}{k!4^k}
  \Bigl[
    (\triangle\psi)      \cdot \frac{\delta}{\delta \psi}
  - (\triangle\psi)^\ast \cdot \frac{\delta}{\delta \psi^\ast}
  \Bigr]^k \psi_n \ .
  \label{3-9}
\end{multline}
The first-term ($k=1$) of the series operates $\psi_n$ as
\begin{equation}
  \Bigl[
    (\triangle\psi)      \cdot \frac{\delta}{\delta \psi}
  - (\triangle\psi)^\ast \cdot \frac{\delta}{\delta \psi^\ast}
  \Bigr] \psi_n 
  =
  \triangle\psi_n \ .
  \label{3-10}
\end{equation}
The second-term ($k=2$) operates as
\begin{multline}
  \Bigl[
    (\triangle\psi)      \cdot \frac{\delta}{\delta \psi}
  - (\triangle\psi)^\ast \cdot \frac{\delta}{\delta \psi^\ast}
  \Bigr]^2 \psi_n  \\%
  =
  \Bigl[
    (\triangle\psi)      \cdot \frac{\delta}{\delta \psi}
  - (\triangle\psi)^\ast \cdot \frac{\delta}{\delta \psi^\ast}
  \Bigr] \triangle\psi_n \\
  =
  (\triangle\psi)      \cdot \frac{\delta \triangle\psi_n}{\delta \psi}
  =
  \triangle \frac{\delta \psi_n}{\delta \psi} \cdot (\triangle\psi)
  =
  \triangle \triangle\psi_n \ .
  \label{3-11}
\end{multline}
Generally,
\begin{equation}
  \Bigl[
    (\triangle\psi)      \cdot \frac{\delta}{\delta \psi}
  - (\triangle\psi)^\ast \cdot \frac{\delta}{\delta \psi^\ast}
  \Bigr]^k \psi_n
  =
  \triangle^k \psi_n \ .
  \label{3-12}
\end{equation}
Thus, we obtain the following identity:
%
\begin{multline}%
  \exp \frac{{\rm i}\Delta{t}}{4} \Bigl[
    (\triangle\psi)      \cdot \frac{\delta}{\delta \psi}
  - (\triangle\psi)^\ast \cdot \frac{\delta}{\delta \psi^\ast}
  \Bigr] \psi_n \\%
  =
  \exp \Bigl[ \frac{{\rm i}\Delta{t}}{4} \triangle \Bigr] \psi_n \ .
  \label{3-13}
\end{multline}%

Similarly, we expand the exponential operator which contains
the effective potential appearing in Eq.~(\ref{3-8}) as
\begin{multline}
  \exp \frac{\Delta{t}}{{\rm i}} \Bigl[
    (V[\rho,t]\psi)      \cdot \frac{\delta}{\delta \psi}
  - (V[\rho,t]\psi)^\ast \cdot \frac{\delta}{\delta \psi^\ast}
  \Bigr] \psi_n = \\
  \sum_{k=0}^{\infty} \frac{(\Delta{t})^k}{k!{\rm i}^k}
  \Bigl[
    (V[\rho,t]\psi)      \cdot \frac{\delta}{\delta \psi}
  - (V[\rho,t]\psi)^\ast \cdot \frac{\delta}{\delta \psi^\ast}
  \Bigr]^k \psi_n \ .
  \label{3-14}
\end{multline}
The first-term ($k=1$) of the series operates $\psi_n$ as
\begin{equation}
  \Bigl[
    (V[\rho,t]\psi)      \cdot \frac{\delta}{\delta \psi}
  - (V[\rho,t]\psi)^\ast \cdot \frac{\delta}{\delta \psi^\ast}
  \Bigr] \psi_n
=
  V[\rho,t] \psi_n \ .
  \label{3-15}
\end{equation}
The second-term ($k=2$) operates as
\begin{multline}
  \Bigl[
    (V[\rho,t]\psi)      \cdot \frac{\delta}{\delta \psi}
  - (V[\rho,t]\psi)^\ast \cdot \frac{\delta}{\delta \psi^\ast}
  \Bigr] V[\rho,t]\psi_n \\
=
  V[\rho,t] V[\rho,t] \psi_n
+ \Bigl(
    (V[\rho,t]\psi)      \cdot
    \frac{\delta V[\rho,t]}{\delta \psi}
  \Bigr) \psi_n \\%
- \Bigl(
    (V[\rho,t]\psi)^\ast \cdot
    \frac{\delta V[\rho,t]}{\delta \psi^\ast}
  \Bigr) \psi_n \\
=
  V[\rho,t] V[\rho,t] \psi_n
+ \Bigl(
    (V[\rho,t]\psi)      \cdot  \psi^\ast\cdot
    \frac{\delta V[\rho,t]}{\delta \rho}
  \Bigr) \psi_n \\%
- \Bigl(
    (V[\rho,t]\psi)^\ast \cdot  \psi\cdot
    \frac{\delta V[\rho,t]}{\delta \rho} 
  \Bigr) \psi_n \\
=
  V[\rho,t] V[\rho,t] \psi_n \ .
  \label{3-16}
\end{multline}
Thus, we obtain the following identity:
%
\begin{multline}%
  \exp \frac{\Delta{t}}{{\rm i}} \Bigl[
    (V[\rho,t]\psi)      \cdot \frac{\delta}{\delta \psi}
  - (V[\rho,t]\psi)^\ast \cdot \frac{\delta}{\delta \psi^\ast}
  \Bigr] \psi_n \\%
=
  \exp \Bigl[ \frac{\Delta{t}}{{\rm i}} V[\rho,t] \Bigr] \psi_n \ .
  \label{3-17}
\end{multline}%

Substituting Eq.~(\ref{3-13}),(\ref{3-17}) into Eq.~(\ref{3-8}), we obtain,
%
\begin{multline}%
  \psi_n(t+\Delta{t})
\simeq
  \exp{\Bigl[ \frac{\Delta{t}}{2} \frac{\partial}{\partial t_{\rm ex}} \Bigr]}
  \exp{\Bigl[ \frac{{\rm i}\Delta{t}}{2} \frac{\triangle}{2}  \Bigr]} \\%
  \exp{\Bigl[ \frac{\Delta{t}}{{\rm i}} V[\rho,t]            \Bigr]} 
  \exp{\Bigl[ \frac{{\rm i}\Delta{t}}{2} \frac{\triangle}{2}  \Bigr]}
  \exp{\Bigl[ \frac{\Delta{t}}{2} \frac{\partial}{\partial t_{\rm ex}} \Bigr]}
  \psi_n(t) \ .
  \label{3-18}
\end{multline}%

By the way, $V_{\rm int}[\rho]$ does not depend on time explicitly,
because the density $\rho$ does not depend on time explicitly as shown in
Eq.~(\ref{3-6}). Meanwhile, $V_{\rm ext}(t)$ does depend on time explicitly,
\begin{equation}
  \frac{\partial V_{\rm int}[\rho]}{\partial t_{\rm ex}} = 0 \ ,
\qquad
  \frac{\partial V_{\rm ext}(t)}{\partial t_{\rm ex}} \neq 0 \ .
  \label{3-19}
\end{equation}

Therefore, the exponential of the explicit-time-derivative operator appearing
in Eq.~(\ref{3-18}) affects only the external time-dependent potential
$V_{\rm ext}(t)$ as
\begin{equation}
  \exp{\Bigl[ \frac{\Delta{t}}{2} \frac{\partial}{\partial t_{\rm ex}} \Bigr]}
  V_{\rm ext}(t) 
=
  V_{\rm ext}(t+\frac{\Delta{t}}{2}) 
  \label{3-20}
\end{equation}

As a result, we obtain the desired formula:
%
\begin{multline}%
  \psi_n(t+\Delta{t})
\simeq
  \exp{\Bigl[ \frac{{\rm i}\Delta{t}}{2} \frac{\triangle}{2}         \Bigr]} \\%
  \exp{\Bigl[ \frac{\Delta{t}}{{\rm i}} \Bigl(
    V_{\rm int}[\rho^\prime] + V_{\rm ext}(t+\frac{\Delta{t}}{2})
   \Bigr) \Bigr]} \\%
  \exp{\Bigl[ \frac{{\rm i}\Delta{t}}{2} \frac{\triangle}{2}         \Bigr]}
  \psi_n(t) \ .
  \label{3-21}
\end{multline}%
%
Here, $V_{\rm ext}(t+\Delta{t}/2)$ is the external force in the middle of the
steps. Meanwhile, $\rho^\prime$ in $V_{\rm int}[\rho^\prime]$ is not the density in the middle of the steps, but it is the density after the preceding operation, namely
\begin{equation}
  \rho^\prime({\bf r})
=
  \sum_{n=1}^N \Bigl| 
    \exp{\Bigl[ \frac{{\rm i}\Delta{t}}{2} \frac{\triangle}{2}         \Bigr]}
    \psi_n({\bf r},t)
  \Bigr|^2 \ .
  \label{3-22}
\end{equation}
Therefore, the formula~(\ref{3-21}) can be explicitly computed without employing
any SCF loops.

The present non-SCF formula~(\ref{3-21}) is quite similar with the conventional
non-SCF formula~(\ref{2-4}) and the conventional SCF formula~(\ref{2-6}).
However, in this paper, we have derived the formula based on the strict
solution~(\ref{3-7}) by considering the time-dependence of the Hamiltonian,
while the conventional non-SCF formula did not consider the time-dependence.
We can easily show that the present non-SCF formula is as accurate as the
conventional SCF formula by associating $\rho^\prime$ with
$\rho(t+\Delta{t})$ as,
\begin{align}
  \rho^\prime
&=
  \sum_{n=1}^N
    |\psi_n(t)|^2
  +{\rm i}\frac{\Delta{t}}{2} \Bigl(
     \psi_n^\ast \frac{\triangle}{2} \psi_n
   - \psi_n      \frac{\triangle}{2} \psi_n^\ast
  \Bigr)_{t} + O(\Delta{t}^2)  \notag\\
&=
  \sum_{n=1}^N
    |\psi_n(t)|^2
  +\frac{\Delta{t}}{2} \Bigl(
     \psi_n^\ast \frac{\partial \psi_n}{\partial t}
   - \psi_n      \frac{\partial \psi_n^\ast}{\partial t}
  \Bigr)_{t} + O(\Delta{t}^2)  \notag\\
&=
  \rho(t) + \frac{\Delta{t}}{2} \frac{\partial \rho}{\partial t}\Bigm|_{t}
 + O(\Delta{t}^2)   \notag\\
&=
  \rho(t+\frac{\Delta{t}}{2}) + O(\Delta{t}^2)  \ .
  \label{3-23}
\end{align}
Therefore, both the non-SCF formula and the SCF formula
are correct up to the second-order of $\Delta{t}$.

\section{Computational technique}

Computational techniques previously developed by us for the one-electron
TD-Schr\"odinger equation\cite{Watanabe2000a,Watanabe2000b} are also
beneficial for formula~(\ref{3-21}). We discretize the wave functions
in real space, and use the finite element method for spatial derivatives.
The only difference in the scheme for the TD-KS equation and TD-Schr\"odinger
equation is the exponential of the effective potential:
\begin{equation}
  \psi_n^\prime({\bf r})
  =
  \exp{\Bigl[ \frac{\Delta{t}}{{\rm i}} V_{\rm int}[\rho] \Bigr]}\,
  \psi_n({\bf r})\ .
  \label{4-1}
\end{equation}
By this operation, the phase of the wave functions is altered at each
point, but the density $\rho({\bf r})$ is not altered. Therefore, we take the
value of $V_{\rm int}[\rho]({\bf r})$ as a constant during the computation,
which is calculated just before the computation.

It is quite easy to improve the accuracy of formula~(\ref{3-21}) to 
the fourth order. The fourth-order accurate formula is given by
Suzuki's exponential product theory\cite{Suzuki1990} as
\begin{multline}
  \psi_n(t+\Delta{t})
\simeq
  S_2(s\Delta{t};t+(1-s)\Delta{t}) \\
  S_2(s\Delta{t};t+(1-2s)\Delta{t}) \,
  S_2((1-4s)\Delta{t};t+2s\Delta{t}) \\
  S_2(s\Delta{t};t+s\Delta{t}) \,
  S_2(s\Delta{t};t)\, \psi_n(t) \ .
  \label{4-2}
\end{multline}
Here, $s$ and $S_2(\Delta{t};t)$ are given as
\begin{equation}
 s = 1/(4-\sqrt[3]{4})
 \label{4-3}
\end{equation}
\begin{multline}%
  S_2(\Delta{t};t)
=
  \exp{\Bigl[ \frac{{\rm i}\Delta{t}}{2} \frac{\triangle}{2} \Bigr]} \\%
  \exp{\Bigl[ \frac{\Delta{t}}{{\rm i}} V[\rho^\prime,t] \Bigr]}
  \exp{\Bigl[ \frac{{\rm i}\Delta{t}}{2} \frac{\triangle}{2} \Bigr]}\ ,
  \label{4-4}
\end{multline}%
%
where, $\rho^\prime$ is the density after the preceding operations.

\section{Example}

In this section, we perform a simple simulation to verify
the efficiency and accuracy of the present method.
The model system we use here is a one-dimensional isolated system
in which two electrons interact by a delta-function interaction under an oscillating
electric field.
The two-body wave function $\Psi(x_1,x_2;t)$ in this system
obeys the following TD-Schr\"odinger equation,
\begin{multline}
  {\rm i} \frac{\partial }{\partial t} \Psi(x_1,x_2;t)
=
  \Bigl[
    - \frac{1}{2} \frac{\partial^2}{\partial x_1^2}
    - \frac{1}{2} \frac{\partial^2}{\partial x_2^2}
    + \alpha \delta(x_1-x_2) \\%
    +  (x_1+x_2) E_o \sin(\omega_o t)
  \Bigr] \Psi(x_1,x_2;t) \ ,
  \label{5-1}
\end{multline}
where $\alpha$ is the coupling constant of the interaction,
and $E_o$ is an external electric field to perturb this system.

We suppose that $\Psi(x_1,x_2;t)$ is expressed by
a common one-electron orbital wave function $\psi(x,t)$ as
%
\begin{multline}%
  \Psi(x_1,x_2;t)
=
  \psi(x_1,t) \psi(x_2,t) \\%
  \frac{1}{\sqrt{2}}
  \Bigl(
    \chi(\uparrow,\sigma_1) \chi(\downarrow,\sigma_2)
  - \chi(\downarrow,\sigma_1) \chi(\uparrow,\sigma_2)
  \Bigr) \ .
  \label{5-2}
\end{multline}%
%
Thus, the TD-KS equation is derived exactly,
\begin{gather}
  {\rm i} \frac{\partial }{\partial t} \psi(x,t)
=
  \Bigl[
    - \frac{1}{2} \frac{\partial^2}{\partial x^2}
    + \alpha \rho(x,t)
    + x E_o \sin(\omega_o t)
  \Bigr] \psi(x,t) \ , \notag\\
  \rho(x,t) = |\psi(x,t)|^2 \ .  \label{5-3}
\end{gather}

We use the following parameters for computation:
\begin{center}
\begin{tabular}{llc}\hline
Size of the system    && $L=8.0$ \\
Number of grid points && $N_p=64$ \\
Mutual interaction    && $\alpha=0.5$ \\
External force        && $E_o=1/64$ \\
Frequency             && $\omega_o=1/8$ \\
Small time slice      && $\Delta{t}=1/16$ \\
Total time steps      && $N_{t}=256{\rm k}$ \\\hline
\end{tabular}
\end{center}

First, we compute the lowest eigen state of this system using
the time-independent Kohn-Sham equation:
\begin{equation}
  E \,\psi_o(x)
=
  \Bigl[
    - \frac{1}{2} \frac{\partial^2}{\partial x^2}
    + \alpha \rho(x)
  \Bigr] \psi_o(x) \ .
  \label{5-4}
\end{equation}
We use this state as the initial state.

\begin{center}
\begin{figure}[h]
  \epsfxsize80mm\mbox{\epsfbox{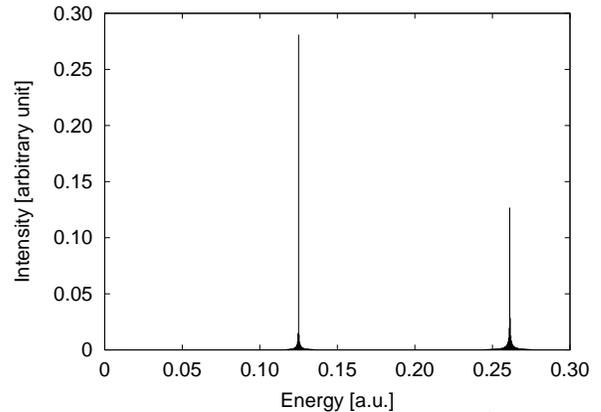}}
  \caption{
    Spectrum of the scattered light. A sharp peak found at $0.125$ is
    corresponding to the Rayleigh scattering. A sharp peak found at $0.261$
    is corresponding to the emission from the first excited state to the
    ground state, this energy includes many-body and non-linear effects.
  }
  \label{fig1}
\end{figure}
\end{center}

Second, we compute the time-evolution using Eq.~(\ref{3-21}).
Third, by Fourier transforming the time-fluctuation of the polarization,
we obtain the spectrum of the scattered light as shown in Figure~\ref{fig1}.

%

The peak appearing in energy $\omega_o=0.125$ comes from the injected light.
The peak appearing in energy $\omega  =0.261$ is expected to be the excitation
energy between the first excited state and the ground state.

We have calculated the excitation energy by certain other methods:
Method (A) solves eigen states by the non-TD-KS equation~(\ref{5-4}),
method (B) modifies the result of (A) by using RPA, and method (C) diagonalizes
the non-TD-Schr\"odinger equation. The results are listed below:

\begin{center}
\begin{tabular}{llr}
\multicolumn{3}{c}{Excitation energies calculated by some methods} \\ \hline
(A) non-TD-KS eq.            && $\omega_{\rm KS} = 0.199$ \\
(B) non-TD-KS eq. with RPA   && $\omega_{\rm RPA}= 0.255$ \\
(C) non-TD-Schr\"odinger eq. && $\omega_{\rm Sch}= 0.260$ \\ \hline
\qquad TD-KS eq.             && $\omega          = 0.261$ \\ \hline
\end{tabular}
\end{center}

We found the peak obtained by the present method, i.e., the TD-KS equation,
reproduces fairly accurately the excitation energy calculated by means of
the exact diagonalization of the non-TD-Schr\"odinger equation.
Namely, by solving the TD-KS equation, dynamical phenomena 
can be described more accurately than using the RPA as far as
the effective Hamiltonian is correct.

Next, to evaluate the error of the method, we estimate the error of the density
 $\rho(x,T)$ at a specified time $T=256$[a.u.].
\begin{equation}
  \text{Error}
=
  \int_{0}^{L}\!\!\! {\rm d}x \,
  \bigl| \rho(x,T) - \rho_{\rm exact}(x,T) \bigr| \ ,
  \label{5-5}
\end{equation}
here the exact value $\rho_{\rm exact}(x,T)$ is prepared in advance by performing
the same simulation on an extremely small time slice  $\Delta{t}=1/256$[a.u.].

\begin{center}
\begin{figure}[h]
  \epsfxsize80mm\mbox{\epsfbox{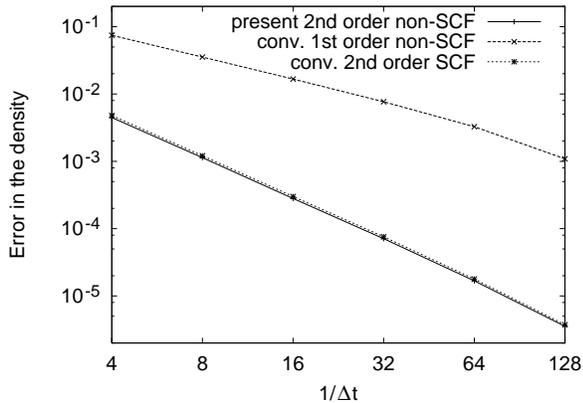}}
  \caption{
    Errors in the density obtained by three methods on some small time slices.
    The conventional non-SCF method is accurate up to the first order of
    $\Delta{t}$, while the present non-SCF method and the conventional SCF
    method are accurate up to the second order of $\Delta{t}$. In this test
    case, the error of the non-SCF method is almost as same as that of the SCF
    method.
  } 
  \label{fig2}
\end{figure}
\end{center}

Figure~\ref{fig2} shows the errors on some time slices obtained by three methods:
the present non-SCF method~(\ref{3-21}),
the conventional non-SCF method~(\ref{2-4}),
and the conventional SCF method~(\ref{2-6}).

%

All methods are accurate enough in this result. However, the conventional
non-SCF method is stable only within a specific short time span:
e.g. $T=512$ [a.u.] for all $\Delta{t}$ in this test.
Meanwhile, the present non-SCF method and the conventional SCF method are stable
even in a long time span: e.g. $T=64M$[a.u.], $\Delta{t}=1/16$[a.u.] in this
test. Therefore, these methods are suitable for long time span simulations.

\begin{center}
\begin{figure}[h]
  \epsfxsize80mm\mbox{\epsfbox{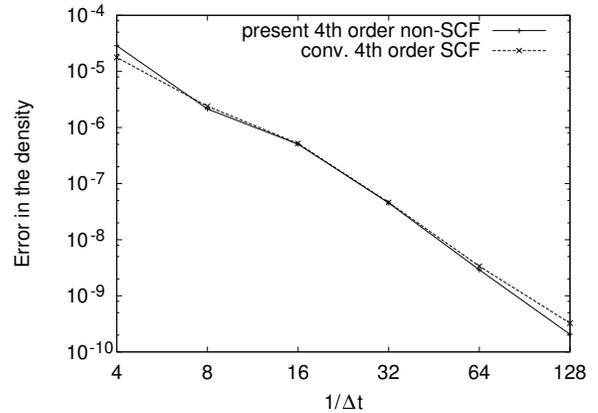}}
  \caption{
    Errors in the density obtained by the fourth-order methods.
    Both errors are roughly proportional to $\Delta{t}^4$, and
    they are much less than those of the second-order methods.
    In this test case, the error of the non-SCF method is almost as same as
    that of the SCF method.
  }
  \label{fig3}
\end{figure}
\end{center}

We have also tested the simulation using the present fourth-order non-SCF
method~(\ref{4-2}) and the fourth-order SCF method proposed
in the literature\cite{Sugino1999}.
Figure~(\ref{fig3}) shows the errors.
Both errors are much less than those of the second-order methods.

%

\section{Conclusion}

We have proved that simulation of the wave function under the TD-KS equation
can be performed by a simple scheme and that there is no need for the use of
SCF-loops to maintain the self-consistency of the effective Hamiltonian.
Our proposed non-SCF method is competitive in accuracy with the SCF
method, and also it is superior in computational efficiency.
We are convinced that our method is helpful for
investigating non-adiabatic and non-linear quantum electrons dynamics.


\end{multicols}

\end{document}